# A Rejoinder on Energy versus Impact Indicators

*Scientometrics* (in press)

Loet Leydesdorff [1] & Tobias Opthof [2,3]

**Abstract**
Citation distributions are so skewed that using the mean or any other central tendency measure is ill-advised. Unlike G. Prathap's scalar measures (Energy, Exergy, and Entropy or *EEE*), the Integrated Impact Indicator (*I3*) is based on non-parametric statistics using the (100) percentiles of the distribution. Observed values can be tested against expected ones; impact can be qualified at the article level and then aggregated.

G. Prathap (2011a) in his Letter applies newly developed scalar measures for bibliometrics (Energy, Exergy, and Entropy; *EEE*) to the data provided in Table 1 of Van Raan (2006, at p. 495). *EEE* operates on averages and ignores the shape of the underlying distributions of citations ("the citation curves"). (Let us note about *EEE* that energy and exergy share dimensionality, but entropy is expressed in Watts/Kelvin. Thus, the expression *Energy – Exergy = Entropy* as suggested by Prathap (2011b) is, in our opinion, invalid the specification of a meta-physical analogon of the "temperature.") Like Prathap (2011b) and following Bornmann & Mutz (2011), Leydesdorff, Bornmann, Mutz, & Opthof (2011) have elaborated the percentile-rank as a scalar sum by using the same dataset that led to the original contention about how citation data should be normalized (Opthof & Leydesdorff, 2010; Van Raan *et al.*, 2010). More recently, Leydesdorff &

[1] Amsterdam School of Communication Research (ASCoR), University of Amsterdam, Kloveniersburgwal 48, 1012 CX Amsterdam, The Netherlands; loet@leydesdorff.net .
[2] Experimental Cardiology Group, Heart Failure Research Center, Academic Medical Center AMC, Meibergdreef 9, 1105 AZ Amsterdam, The Netherlands; t.opthof@inter.nl.net.
[3] Department Of Medical Physiology, University Medical Center Utrecht, The Netherlands

Bornmann (2011) have developed this scalar measure into the Integrated Impact Indicator (*I3*).[4]

The difference between *I3* and *EEE* is that *I3* takes the shapes of the distribution into account and allows for non-parametric significance tests, whereas Prathap's systems view ignores this shape and uses averages on the assumption of the Central Limit Theorem (Glänzel, 2010). However, citation distributions are extremely skewed (Seglen, 1992; 1997; cf. Leydesdorff, 2008) and central tendency statistics give misleading results. Using parametric statistics, one can neither reliably test the significance of observations nor the significance of differences in rankings.

Prathap (2011a) was able to compute using the mean values of JCS (Journal Citation Scores) and FCS (Field Citation Scores) because his concept of entropy is no longer *probabilistic* entropy (cf. Leydesdorff, 1995; Theil, 1972), but thermodynamic entropy (Prathap, 2011b, at p. 523f.). However, the impact of two hits is not their average, but their sum. In the case of collisions, this is the vector sum of the momenta. We agree that in the case of citations one should use a scalar sum.

The scalar sum of citations (that is, total citations) would as yet be insufficiently qualified. The quality along the skewed citation curve must first be normalized in terms of percentiles. Bornmann & Mutz (2011) normalized in terms of six percentile-rank classes, but the more general case is normalization in terms of quantiles as a continuous variable which can thereafter be organized using different evaluation schemes (Leydesdorff *et al.*, 2011; Leydesdorff & Bornmann, 2011a; Rousseau, 2011). Different aggregations are possible because the impacts, once normalized in terms of percentiles, are determined at the paper level. This Integrated Impact Indicator (*I3*) can be formalized as an integration as follows:

$$I3 = \sum_i x_i * f(x_i) \tag{1}$$

---

[4] The software for measuring this indicator is available at http://www.leydesdorff.net/software/i3.

Citations are discrete events and therefore the integral is in this case a step function: using Equation 1, the frequency of papers in each percentile ($x_i$) is multiplied by the percentile of each paper ($f(x_i)$). The resulting scalar ($\Sigma$) of the total impact can then be scaled (*i*) in terms of various evaluation schemes (e.g., quartiles, or the six evaluation categories used in the U.S. *Science & Engineering Indicators* (NSB, 2010) and by Bornmann & Mutz (2011)); (*ii*) tested for their significance against a theoretically specified expectation; (*iii*) expressed as a single number, namely a percentage of total impact contained in the reference set; and (*iv*) used to compare among and between various units of analysis such as journals, countries, institutes, and cities; by aggregating cases in a statistically controllable way (Theil, 1972).

In summary, the discussion over *Rates of Averages* versus *Averages of Rates* (Gingras and Larivière, 2011) has taught us that a rate of averages is merely a quotient number that does not allow for testing, and is mathematically inconsistent (Waltman *et al.*, 2011). The mean observed citation ratio (*MOCR*) should not be divided by the mean expected citation ratio (*RCR* = *MOCR*/*MECR*; Schubert & Braun, 1986; cf. Glänzel *et al.*, 2009, at

p. 182), but observed values can be *tested against* expected values by using appropriate statistics.

Secondly, citation indicators based on averaging skewed distributions—such as Prathap's *EEE* and the new "crown indicator" *MNCS*—are unreliable. For example, Leydesdorff *et al.* (2011) have shown that in the case of seven Principal Investigators at the Academic Medical Center of the University of Amsterdam, the number one ranked PI would fall to fifth position, whereas the sixth-ranked PI would become the highest-ranked author if percentiles or percentile ranks are used.

Thirdly, one should not test sets of documents as independent samples against each other, but as subsets of a reference set (Bornmann *et al.*, 2008): each subset contributes a percentage impact to the set. The reference set allows for normalization and the specification of an expectation. (This specification can further be informed on theoretical grounds.) Using quantiles and percentile ranks, the observed values can be tested against the expected ones using non-parametric statistics.

Furthermore, and not specific as criticism of *EEE*, field delineations do not have to be based on *ex ante* classification schemes such as the ISI Subject Categories. Hitherto, journal classifications have been unprecise and unreliable (Boyack & Klavans, 2011; Leydesdorff, 2006; Pudovkin & Garfield, 2002; Rafols *et al.*, 2009). Fractional attribution of citations in the citing documents, however, can be used for normalization of differences in citation potentials (Garfield, 1979) reflecting differences in citation behavior at the level of individual papers (Leydesdorff & Bornmann, 2011b; Leydesdorff & Opthof, 2010; Moed, 2010).

Given these recent improvements in citation normalization—such as the use of paper-based measures both cited and citing—the theoretical question remains whether citations can be used as indicators of scientific quality, and if so, when? (Amsterdamska & Leydesdorff, 1989; Bornmann *et al*., 2008; Garfield, 1979; Leydesdorff, 1998; Leydesdorff & Amsterdamska, 1990). Opthof & Leydesdorff (2010) opened this discussion by asking whether citation analysis enables us to legitimate the strategic selection of "excellent" as as against merely "good" research?